# Double MRT Thermal Lattice Boltzmann Method for Simulating Natural Convection of Low Prandtl Number Fluids

Zheng Li[a,b], Mo Yang[a] and Yuwen Zhang[b,1]
[a] School of Energy and Power Engineering, University of Shanghai for Science and Technology, Shanghai 200093, China
[b] Department of Mechanical and Aerospace Engineering, University of Missouri, Columbia, MO 65211, USA

**ABSTRACT**
**Purpose:** The purposes of this paper are testing an efficiency algorithm based on LBM and using it to analyze two-dimensional natural convection with low Prandtl number.
**Design/methodology/approach:** Steady state or oscillatory results are obtained using double multiple-relaxation-time thermal lattice Boltzmann method. The velocity and temperature fields are solved using D2Q9 and D2Q5 models, respectively.
**Findings:** With different Rayleigh number, the tested natural convection can either achieve to steady state or oscillatory. With fixed Rayleigh number, lower Prandtl number leads to a weaker convection effect, longer oscillation period and higher oscillation amplitude for the cases reaching oscillatory solutions. At fixed Prandtl number, higher Rayleigh number leads to a more notable convection effect and longer oscillation period.
**Originality/value:** Double multiple-relaxation-time thermal lattice Boltzmann method is applied to simulate the low Prandtl number (0.001 – 0.01) fluid natural convection. Rayleigh number and Prandtl number effects are also investigated when the natural convection results oscillate.

**NOMENCLATURE**
$c$ lattice speed
$c_p$ specific heat (J/kgK)
$c_s$ sound speed
$\mathbf{e}_i$ particle speed
$f_i$ density distribution
$F_i$ body force
$Fo$ Fourier number
$g$ gravity acceleration $(m/s^2)$
$G$ effective gravitational acceleration $(m/s^2)$
$g_i$ energy distribution
$k$ thermal conductivity (W/m k)
$M$ transform matrix for density distribution
$m_i$ moment function for density distribution
$Ma$ Mach number

---

[1] Corresponding author. Email: zhangyu@missouri.edu. Tel: 001-573-884-6936. Fax: 001-573-884-5090

$N$ transform matrix for density distribution

$n_i$ moment function for energy distribution

$p$ pressure (Pa)

$P$ non-dimensional pressure

$Pr$ Prandtl number

$Q$ collision matrix for energy distribution

$Ra$ Rayleigh number

$s$ relaxation time in density distribution

$S$ collision matrix for density distribution

$t$ time (s)

$T$ temperature $(K)$

$u$ velocity in x-direction (m/s)

$u_i$ particle speed in energy distribution

$U$ non-dimensional velocity in x-direction

$v$ velocity in y-direction (m/s)

$V$ non-dimensional velocity in y-direction

$\mathbf{V}$ velocity

$\alpha$ thermal diffusivity $(m^2/s)$

$\beta$ thermal expansion (K$^{-1}$)

$\Delta t$ time step (s)

$\theta$ non-dimensional temperature

$\mu$ viscosity (Kg/ms)

$\rho$ Density (kg/m$^3$)

$\sigma$ relaxation time in energy distribution

$\tau$ non-dimensional time

$\nu$ kinematic viscosity $(m^2/s)$

# 1. Introduction

Lattice Boltzmann method (LBM) has been developed into a promising numerical method in the last two decades. It can be used to solve different fluid flow problems, such as incompressible fluid flow (Guo and Zhao, 2002), compressible fluid flow (Kataoka and Tsutahara, 2004) and multiphase fluid flow (Luo, 2000). Instead of solving the macroscopic continuum and momentum equations as the traditional computational fluid dynamics (CFD), the LBM is based on solving the discrete Boltzmann equation in statistical physics via two basic steps: collision step and streaming step (Succi, 2001). There are different LBM models for fluid flow problems. Lattice Bhatnagar-Gross-Krook (LBGK) simplifies the collision term with one relaxation time (Chen and Chen, 1991; Chen and Doolen, 1998). Based on LBGK model, Li et al. (2014a) and Li et al. (2014b) use combined LBM and finite volume method to solve lid driven flow and natural convection, respectively. Although it is widely used, LBGK is limited by the numerical instability (Lallemand and Luo, 2000). To overcome this limitation, entropy LBM (ELBM) (Chikatamarla et al., 2006; Chikatamarla and Karlin, 2006), two-relaxation-time model (TRT) (Ginzburg, 2005; Ginzburg and d'Humieres, 2007) and multiple relaxation time model (MRT) (Lallemand and Luo, 2000; Lallemand and Luo, 2003) have been proposed. The difference among these models lies in the ways to simplify the collision term while their streaming steps are the same. Luo et al. (2011) compared these models by using them to solve the lid driven flow problem. It was concluded that the MRT was preferred due to its advantages in accuracy and numerical stability.

The fluid flow problem with heat transfer also can be solved using LBM. Multispeed approach (MS), hybrid method and double distribution functions (DDF) are the common thermal LBM models. The MS approach obtains the temperature field by adding more discrete velocities to the density distribution (Chen et al., 1994). It is limited by numerical instability and narrow range of temperature variation (Guo et al., 2002). Hybrid method uses LBM to solve the velocity field and employs other numerical method, such as finite volume method (FVM), to obtain the temperature field. Li et al. (2014c) and Li et al. (2014d) used hybrid LBM-FVM to solve the natural convection and melting problems. The DDF employs two independent distributions to analyze the momentum and energy equations (He et al., 1998). It has been applied to solve different kinds of heat transfer problems (Peng et al., 2003; Huber et al., 2008; Gao and Chen, 2011).

Most of the reported DDF results are based on LBGK. Several double MRT models are proposed for the fluid flow and heat transfer problem in the recent years. Mezrhab et al. (2010) used double MRT thermal LBM for simulating air convective flow in a cavity respectively for Rayleigh number up to $1 \times 10^8$. Wang et al. (2013) discussed the convective flow for Prandtl number of 0.71 and 7.0 with the double MRT thermal LBM. On the other hand, convective flow of low-Prandtl number fluid is important in many industry applications (Li et al., 2015). Low Prandtl number convection problem involves highly nonlinear fluid dynamics. It has more possibility to reach oscillatory results. Li et al. (2015) discussed low Prandtl number melting problem with double LBGK model. Kosec and Sarler (2013) reported the solution of a low Prandtl number natural convection benchmark problem. Kosec and Sarler (2014) solved low Prandtl number solidification problems using a meshless method. The objective of this paper is to employ the double MRT (Wang et al. 2013) to analyze the low Prandtl number natural convection problems.

## 2. Problem Statement

As shown in Fig. 1, the considered natural convection domain is a square-shaped cavity filled with incompressible fluid. The cavity height and width are $H$. The left boundary is kept a constant temperature $T_h$ and the right boundary has a lower constant temperature of $T_c$. Meanwhile, the top and bottom boundaries are adiabatic. Non-slip boundary condition is applied to all boundaries. Applying Boussinesq assumption, the problem can be described by the following governing equations:

$$\frac{\partial u}{\partial x} + \frac{\partial v}{\partial y} = 0 \tag{1}$$

$$\rho \left[ \frac{\partial u}{\partial t} + u \frac{\partial u}{\partial x} + v \frac{\partial u}{\partial y} \right] = -\frac{\partial p}{\partial x} + \mu \left( \frac{\partial^2 u}{\partial x^2} + \frac{\partial^2 u}{\partial y^2} \right) \tag{2}$$

$$\rho \left[ \frac{\partial v}{\partial t} + u \frac{\partial v}{\partial x} + v \frac{\partial v}{\partial y} \right] = -\frac{\partial p}{\partial y} + \mu \left( \frac{\partial^2 v}{\partial x^2} + \frac{\partial^2 v}{\partial y^2} \right) + \rho g \beta (T - T_c) \tag{3}$$

$$(\rho c_p) \left[ \frac{\partial T}{\partial t} + u \frac{\partial T}{\partial x} + v \frac{\partial T}{\partial y} \right] = k \left( \frac{\partial^2 T}{\partial x^2} + \frac{\partial^2 T}{\partial y^2} \right) \tag{4}$$

Equations (1) – (4) are subject to the following boundary conditions:

$$x = 0, u = 0, v = 0, T = T_h \tag{5}$$

$$x = H, u = 0, v = 0, T = T_c \tag{6}$$

$$y = 0, u = 0, v = 0, \partial T / \partial y = 0 \tag{7}$$

$$y = H, u = 0, v = 0, \partial T / \partial y = 0 \tag{8}$$

Defining the following non-dimensional variables

$$\begin{cases} X = \dfrac{x}{H}, Y = \dfrac{y}{H}, u_c = \sqrt{g\beta(T_h - T_c)H}, Ma = \dfrac{u_c}{c_s}, U = \dfrac{u}{\sqrt{3}c_s}, V = \dfrac{v}{\sqrt{3}c_s}, \\ \tau = \dfrac{t \cdot \sqrt{3}c_s}{H}, \theta = \dfrac{T - T_c}{T_h - T_c}, P = \dfrac{p}{3\rho c_s^2}, Pr = \dfrac{\nu}{\alpha}, Ra = \dfrac{g\beta(T_h - T_c)H^3 Pr}{\nu^2} \end{cases} \tag{9}$$

where $c_s$, $\nu$ and $\alpha$ are the speed of sound, kinematic viscosity and thermal diffusivity resepectively. Equations. (1) to (8) can be nondimensionalized to:

$$\frac{\partial U}{\partial X} + \frac{\partial V}{\partial Y} = 0 \tag{10}$$

$$\frac{\partial U}{\partial \tau} + U \frac{\partial U}{\partial X} + V \frac{\partial U}{\partial Y} = -\frac{\partial P}{\partial X} + Ma\sqrt{\frac{Pr}{3Ra}} \left( \frac{\partial^2 U}{\partial X^2} + \frac{\partial^2 U}{\partial Y^2} \right) \tag{11}$$

$$\frac{\partial V}{\partial \tau} + U \frac{\partial V}{\partial X} + V \frac{\partial V}{\partial Y} = -\frac{\partial P}{\partial Y} + Ma\sqrt{\frac{Pr}{3Ra}} \left( \frac{\partial^2 V}{\partial X^2} + \frac{\partial^2 V}{\partial Y^2} \right) + \frac{Ma^2 \theta}{3} \tag{12}$$

$$\frac{\partial \theta}{\partial \tau} + U \frac{\partial \theta}{\partial X} + V \frac{\partial \theta}{\partial Y} = Ma\sqrt{\frac{1}{3Ra \cdot Pr}} \left( \frac{\partial^2 \theta}{\partial X^2} + \frac{\partial^2 \theta}{\partial Y^2} \right) \tag{13}$$

$$X = 0, U = 0, V = 0, \theta = 1 \tag{14}$$

$$X = 1, U = 0, V = 0, \theta = 0 \tag{15}$$

$$Y = 0, U = 0, V = 0, \partial \theta / \partial Y = 0 \tag{16}$$

$$Y = 1, U = 0, U = 0, \partial \theta / \partial Y = 0 \tag{17}$$

Heat transfer is evaluated based on Nusselt number $Nu$, which is the ratio of convection to conduction heat transfer across the boundary:

$$Nu = \frac{h}{k/H}\bigg|_{x=0} = -\frac{\partial \theta}{\partial X}\bigg|_{x=0} \tag{18}$$

where $h$ is the convective heat transfer coefficient. Average Nusselt number can be obtained by the following equation:

$$Nu_{avg} = \int_0^1 Nu \, dY \tag{19}$$

It represents the average heat transfer rate through the left heat wall. Meanwhile, the Fourier number $Fo$ which is defined as

$$Fo = \alpha t / H^2 \tag{20}$$

is another non-dimensional time parameter. It has a relation to the non-dimensional time $\tau$ as following:

$$Fo = \tau \cdot Ma / \sqrt{3Ra \cdot Pr} \tag{21}$$

## 3. Double MRT thermal lattice Boltzmann model

Double MRT thermal LBM model (Wang et al., 2013) is selected to solve the natural convection problem. D2Q9-MRT is applied to analyze the velocity flied and the temperature field is solved by D2Q5-MRT.

### 3.1 D2Q9-MRT for fluid flow

Lattice Boltzmann equation can describe the statistical behavior of a fluid flow.

$$f(\mathbf{r}+\mathbf{e}\Delta t, t+\Delta t)-f(\mathbf{r},t)=\Omega+F \tag{22}$$

where $f$ is the density distribution, $\Delta t$ is the time step, $\Omega$ is the collision term and $F$ is the body force. D2Q9 model is preferred for the velocity field because the problem in consideration is two-dimensional. Each computing nodes has nine local particle velocities shown in Fig. 2. These velocities are given by:

$$\mathbf{e}_i = \begin{cases} (0,0) & i=1 \\ c(-\cos\dfrac{i\pi}{2}, -\sin\dfrac{i\pi}{2}) & i=2,3,4,5 \\ \sqrt{2}c(-\cos\dfrac{(2i+1)\pi}{4}, -\sin\dfrac{(2i+1)\pi}{4}) & i=6,7,8,9 \end{cases} \tag{23}$$

where $c$ is the lattice speed and relates to the sound speed $c_s$ as:

$$c^2 = 3c_s^2 \tag{24}$$

Then Eq. (22) can be expressed as:

$$f_i(\mathbf{r}+\mathbf{e}_i\Delta t, t+\Delta t)-f_i(\mathbf{r},t)=\Omega_i+F_i, \quad i=1,2,\ldots 9 \tag{25}$$

The force term in the equation above can be obtained as:

$$F_i = \Delta t \mathbf{G} \cdot \frac{(\mathbf{e}_i - \mathbf{V})}{p} f_i^{eq} \tag{26}$$

where $\mathbf{G}$ is the effective gravitational force:

$$\mathbf{G} = -\beta(\theta-\theta_l)\mathbf{g} \tag{27}$$

To satisfy the continuum and momentum conservations, the collision term in MRT is:

$$\Omega_i = -M^{-1} \cdot S \cdot \left[ m_i(\mathbf{r},t) - m_i^{eq}(\mathbf{r},t) \right], \quad i=1,2,\ldots 9 \tag{28}$$

where $m_i(\mathbf{r},t)$ and $m_i^{eq}(\mathbf{r},t)$ are moments and their equilibrium functions; $M$ and $S$ are the transform matrix and collision matrix respectively (Mezrhab et al., 2010).

For the D2Q9 model the nine macroscopic moments are:

$$\begin{aligned}\mathbf{m} &= (m_1, m_2, m_3, m_4, m_5, m_6, m_7, m_8, m_9)^T \\ &= (\rho, j_x, j_y, e, p_{xx}, p_{xy}, q_x, q_y, \varepsilon)^T \end{aligned} \tag{29}$$

Consequently, the transform matrix $M$ and is:

$$M = \begin{pmatrix} 1 & 1 & 1 & 1 & 1 & 1 & 1 & 1 & 1 \\ 0 & 1 & 0 & -1 & 0 & 1 & -1 & -1 & 1 \\ 0 & 0 & 1 & 0 & -1 & 1 & 1 & -1 & -1 \\ -4 & -1 & -1 & -1 & -1 & 2 & 2 & 2 & 2 \\ 0 & 1 & -1 & 1 & -1 & 0 & 0 & 0 & 0 \\ 0 & 0 & 0 & 0 & 0 & 1 & -1 & 1 & -1 \\ 0 & -2 & 0 & 2 & 0 & 1 & -1 & -1 & 1 \\ 0 & 0 & -2 & 0 & 2 & 1 & 1 & -1 & -1 \\ 4 & -2 & -2 & -2 & -2 & 1 & 1 & 1 & 1 \end{pmatrix} \tag{30}$$

Correspondingly, the collision matrix $S$ is:

$$S = diag(0, 1, 1, s_e, s_v, s_v, s_q, s_q, s_\varepsilon) \tag{31}$$

In the present simulations, the unknown parameters in Eq. (31) are defined as following:

$$\begin{cases} s_e = s_\varepsilon = s_\nu = \dfrac{2}{6\nu+1} \\ s_q = 8\dfrac{(2-s_\nu)}{(8-s_\nu)} \end{cases} \quad (32)$$

Then $\rho$, $j_x$ and $j_y$ are related with density distribution $f_i$ by the following equation:

$$\begin{cases} \rho = \delta\rho + \rho_0, \rho_0 = 1, \delta\rho = \sum_{i=1}^{9} f_i \\ \boldsymbol{j} = \rho_0 \boldsymbol{u} = \rho_0(u,v), \rho_0 \boldsymbol{u} = \sum_{i=1}^{9} f_i \boldsymbol{e}_i \end{cases} \quad (33)$$

Adding $\rho_0$ can reduce round-off errors in the simulation process (Wang et al. 2013). Accordingly, the equilibrium functions $m_i^{eq}$ are

$$\begin{cases} m_1^{eq} = \delta\rho, m_2^{eq} = \rho_0 u, m_3^{eq} = \rho_0 v, m_4^{eq} = -2\delta\rho + \rho_0 \boldsymbol{u}\cdot\boldsymbol{u}, m_5^{eq} = \rho_0(u^2 - v^2) \\ m_6^{eq} = \rho_0 uv, m_7^{eq} = -\rho_0 u, m_8^{eq} = -\rho_0 v, m_9^{eq} = \delta\rho - 3\rho_0 \boldsymbol{u}\cdot\boldsymbol{u} \end{cases} \quad (34)$$

Bounce-back scheme is employed for the no-slip boundary conditions (Latt and Chopard, 2008).

## 3.2 D2Q5-MRT for heat transfer

The D2Q5 model is used to obtain the temperature field. Each computing node has five discrete velocities shown is Fig. 3:

$$u_i = \begin{cases} (0,0) & i=1 \\ c(-\cos\dfrac{i\pi}{2}, -\sin\dfrac{i\pi}{2}) & i=2,3,4,5 \end{cases} \quad (35)$$

Similar to the density distribution, the energy distribution $g_i$ can be obtained by:

$$g_i(\boldsymbol{r}+\boldsymbol{u}_i\Delta t, t+\Delta t) - g_i(\boldsymbol{r},t) = -N^{-1}\cdot Q\cdot\left[n_i(\boldsymbol{r},t) - n_i^{eq}(\boldsymbol{r},t)\right], \; i=1,2,...5 \quad (36)$$

where $n_i^{eq}(\boldsymbol{r},t)$ are the equilibrium functions for $n_i(\boldsymbol{r},t)$, $N$ and $Q$ are the transform matrix and collision matrix for the energy distribution (Wang et al. 2013):

$$N = \begin{pmatrix} 1 & 1 & 1 & 1 & 1 \\ 0 & 1 & 0 & -1 & 0 \\ 0 & 0 & 1 & 0 & -1 \\ -4 & 1 & 1 & 1 & 1 \\ 0 & 1 & -1 & 1 & 1 \end{pmatrix} \quad (37)$$

$$Q = diag(0, \sigma_k, \sigma_k, \sigma_e, \sigma_\nu) \quad (38)$$

where $\left(\dfrac{1}{\sigma_\nu} - \dfrac{1}{2}\right)\left(\dfrac{1}{\sigma_k} - \dfrac{1}{2}\right) = \dfrac{1}{6}$ \quad (39)

The unknown parameters in Eq. (38) are defined as following (Wang et al. 2013):

$$\begin{cases} \dfrac{1}{\sigma_e} - \dfrac{1}{2} = \dfrac{1}{\sigma_\nu} - \dfrac{1}{2} = \dfrac{\sqrt{3}}{3} \\ \dfrac{1}{\sigma_k} - \dfrac{1}{2} = \dfrac{\sqrt{3}}{6} \end{cases} \quad (40)$$

The temperature at each computing node can be obtained as:

$$\theta = \sum_{i=1}^{5} g_i \tag{41}$$

Then the equilibrium functions $n_i^{eq}(\mathbf{r},t)$ are:

$$n_1^{eq} = T, n_2^{eq} = uT, n_3^{eq} = vT, n_4^{eq} = aT, n_5^{eq} = 0 \tag{42}$$

where $a$ is related to the thermal diffusivity $\alpha$ by:

$$\alpha = \frac{(4+a)}{10}\left(\frac{1}{\sigma_k} - \frac{1}{2}\right) \tag{43}$$

It is necessary that $a < 1$ to avoid instability (Ginzburg, 2012). The natural convection problem under consideration involves two types of boundary conditions: constant temperature and adiabatic. Assuming $x_f$ is a fluid computing node adjacent to the boundary, only one direction energy distribution $g_i(x_f,t)$ among the five directions in D2Q5 model is unknown on each boundary nodes. On its opposite direction, energy distribution $g_{\bar{i}}(x_f,t)$ is known after the streaming process. The top and bottom of the cavity are adiabatic and bounce-back scheme (Wang et al. 2013) is employed to fulfill them:

$$g_i(x_f,t) = g_{\bar{i}}(x_f,t) \tag{44}$$

For the boundary with constant temperature $\theta_w$, the anti-bounce-back boundary condition is employed (Mezrhab et al., 2010).

$$g_i(x_f,t) = 2\sqrt{3}\alpha\theta_w - g_{\bar{i}}(x_f,t) \tag{45}$$

# 4. Results and discussions

The natural convection problem is governed by Rayleigh number and Prandtl number. The Rayleigh number, $Ra$ is depending on the temperature difference $T_h - T_c$, cavity height $H$ and thermal properties of fluid. Meanwhile, the Prandtl number, $Pr$ is a fluid thermal property that varies from $10^{-3}$ (liquid metal) to $10^5$ (functional oil) (Kosec and Sarler, 2013). The objective of this paper is to study natural convection of low $Pr$ (orders of magnitude from $10^{-3}$ to $10^{-2}$) fluid. Four test cases are solved with double MRT model. Their Rayleigh numbers and Prandtl numbers are listed in Table 1.

Natural convection in a square enclosure with $Ra = 10^4$ and $Pr = 0.01$ is considered first (referred to as Case 1 thereafter). Figure 4 shows the variation of the average Nusselt number with time. It can be seen that the average Nusselt number become a constant after Fo = 3; this indicates the heat transfer rate through the left heat wall reaches a fixed value of 1.95 which agrees with that in Kosec and Sarler (2013) well. Figure 5 shows the streamlines for Case 1 that one vortex exists in the center of cavity due to the convection effect; it agrees with that in Kosec and Sarler (2013) as well. This convection effect is also evident in the temperature field shown in Fig. 6. The local Nusselt number along the vertical direction is shown in Fig. 7. The maximum Nusselt number $Nu_{max}$ of 3.02 occurs at the location of $Y_{Nu_{max}} = 0.30$.

Simulation is then carried out for Case 2 that Rayleigh number increases to $5 \times 10^4$ while the Prandtl number is kept at $0.01$. Figure 8 shows the variation of the average Nusselt number with time. It is different from the Case 1 that steady state cannot be reached; After Fo =1, the average Nusselt number oscillates around 2.80, which agrees with that in Kosec and Sarler (2013). The amplitudes and periods of the oscillations turn to be constant (0.01 and 0.092) as the time increasing. The streamlines and temperature fields in one oscillation at different times are presented in Figs. 9 and 10, respectively. Convection effect is more evident than that in Case 1 due to the higher Rayleigh number. A stronger vortex exists in the cavity and changes with time. It leads to temperature fields changing with times. The streamlines difference is more evident than that in temperature fields.

Simulation is now carried out for a lower Prandtl number of 0.005 with Ra = $5 \times 10^4$. Similar to Case 2, it can still reach oscillatory solution. The variation of average Nusselt number with time is shown in Fig. 11. It oscillates around 2.65 after $Fo$ =1.5, and the mean value of the Nusselt number is lower than that of Case 2. In other words, the convection effect weakens as Prandtl number decreases. The amplitude and period of oscillation are 0.05 and 0.093, respectively. Figures 12 and 13 show streamlines and temperature fields in one oscillation at different times. It is clear that lower Prandtl number leads to a lower convection effect, longer oscillation period and higher oscillation amplitude.

To further study the effects of Rayleigh number on the natural convection, another case is studied for Ra = $1 \times 10^5$ and Pr = 0.01 (Case 4). This case also reaches the oscillatory solution. Instead of oscillating round a constant value as that in Cases 2 or 3, average Nusselt number for Case 4 varies round a wave as shown in Fig. 14. The streamlines and temperature fields in one period (0.073) at different times are shown in Figs. 15 and 16, respectively. Comparing with case 2, the convection effect is more notable because of the higher Rayleigh number. And the oscillation period is also longer than that for Case 2.

## Conclusions

Double MRT thermal LBM is applied to simulate the natural convection of fluid with low Prandtl number ($10^{-3}$ – $10^{-2}$). The natural convection can reach to steady state or oscillate, which agree with the reference results well. Therefore, the double MRT thermal LBM is valid for simulation of natural convection of the fluid with low Prandtl numbers. With fixed Rayleigh number, lower Prandtl number leads to a weaker convection effect, longer oscillation period and higher oscillation amplitude for the cases reaching oscillatory solutions. At fixed Prandtl number, higher Rayleigh number leads to a more notable convection effect and longer oscillation period.

**Acknowledgement**

Support for this work by the U.S. National Science Foundation under grant number CBET-1404482, Chinese National Natural Science Foundations under Grants 51129602 and 51276118 are gratefully acknowledged.

## Reference

Chen S., Chen H., Martnez D. and Matthaeus W., Lattice Boltzmann model for simulation of magnetohydrodynamics, Physical Review Letters, 67 (27), pp. 3776-3779, 1991.

Chen S. and Doolen G., Lattice Boltzmann method for fluid flows. Annual Review of Fluid Mechanics, 30, pp.329–364, 1998.

Chen Y., Ohashi H. and Akiyama M., Thermal lattice Bhatnagar–Gross–Krook model without nonlinear deviations in macrodynamic equations, Physics Review E, 50, pp.2776 –2783, 1994.

Chikatamarla S., Ansumali S. and Karlin I., Entropic lattice Boltzmann models for hydrodynamics in three dimensions, Physical Review Letter, 97 (1), pp. 010201, 2006.

Chikatamarla S. and Karlin I., Entropy and Galilean invariance of lattice Boltzmann

theories, Physical Review Letter, 97 (19), pp. 190601, 2006.

Gao D. and Chen Z., Lattice Boltzmann simulation of natural convection dominated melting in a rectangular cavity filled with porous media, International of Journal of Thermal Sciences, 50 (4), pp. 493-501, 2011.

Ginzburg I., Equilibrium-type and link-type lattice Boltzmann models for generic advection and anisotropic-dispersion equation, Advance in Water Resources, 28 (11), pp. 1171-1195, 2005.

Ginzburg I., Truncation errors, exact and heuristic stability analysis of two-relaxation-times lattice Boltzmann schemes for anisotropic advection-diffusion equation, Communications in Computational Physics, 11 (5), pp. 1090-1143, 2012.

Ginzbirg I. and d'Humieres D., Lattice Boltzmann and analytical modeling of flow processes in anisotropic and heterogeneous stratified aquifers, Advance in Water Resources, 30 (11), pp. 2202-2234, 2007.

Guo Z., Shi B. and Zheng C., A coupled lattice BGK model for the Boussinesq equations, International Journal for Numerical Methods in Fluids, 39 (4), pp. 325–342, 2002.

Guo Z. and Zhao T., Lattice Boltzmann model for incompressible flow through porous media, Physical Review E, 66, pp. 036304, 2002.

He X., Chen S. and Doolen G., A novel thermal model for the lattice Boltzmann method in incompressible limit, Journal of Computational Physics, 146 (1), pp. 282–300, 1998.

Huber C., Parmigiani A., Chopard B., Manga M. and Bachmann O., Lattice Boltzmann model for melting with natural convection, International Journal of Heat and Fluid Flow, 29, pp. 1469-1480, 2008.

Kataoka T. and Tsutahara M., Lattice Boltzmann model for the compressible Navier-Stokes equations with flexible specific-heat ratio, Physical Review E, 69, pp. 035701, 2004.

Kosec G. and Sarler B., Solution of a low Prandtl number natural convection benchmark by a local meshless method, International Journal of Numerical Methods for Heat & Fluid flow, 23 (1), pp. 189-204, 2013.

Kosec G., Šarler B. Simulation of macrosegregation with mesosegregates in binary metallic casts by a meshless method, Engineering Analysis with Boundary Elements, 45, pp. 36-44, 2014.

Lallemand P. and Luo L., Theory of the lattice Boltzmann method: dispersion, dissipation, isotropy, Galilean invariance, and stability, Physical Review E, 61 (6), pp. 6546-6562, 2000.

Lallemand P. and Luo L., Theory of the lattice Boltzmann method: Acoustic and thermal properties in two and three dimensions, Physical Review E, 68 (3), pp. 036706, 2003.

Latt J. and Chopard B., Straight velocity boundaries in the lattice Boltzmann method. Physical Review E, 77, pp. 056703, 2008.

Li Z., Yang M. and Zhang Y., Hybrid lattice Boltzmann and finite volume methods for fluid flow problems. International Journal for Multiscale Computational Engineering, 13 (3), pp. 177-192, 2014a.

Li Z., Yang M. and Zhang Y., A coupled lattice Boltzmann and finite volume method for natural convection simulation. International Journal of Heat and Mass Transfer, 70, pp. 864-874, 2014b.

Li Z., Yang M. and Zhang Y., Hybrid lattice Boltzmann and finite volume method for natural convection, Journal of Thermophysics and Heat Transfer, 28 (1), pp. 68-77, 2014c.

Li Z., Yang M., and Zhang Y., A hybrid lattice Boltzmann and finite volume method for melting with natural convection, Numerical Heat Transfer, Part B: Fundamentals, 66 (4), pp. 307-325, 2014d.

Li, Z., Yang, M. and Zhang, Numerical Simulation of Melting Problems Using the Lattice Boltzmann Method with the Interfacial Tracking Method, Numerical Heat Transfer, Part A: Applications, 68 (11), pp. 1175-1197, 2015.

Luo L., Theory of the lattice Boltzmann method: Lattice Boltzmann models for nonideal gases, Physical Review E, 62, pp. 4982, 2000.

Luo L., Liao W., Chen X., Peng Y. and Zhang W., Numerics of the lattice Boltzmann method: Effects of collision models on the lattice Boltzmann simulations, Physical Review E, 83 (5), pp. 056710, 2011.

Mezrhab A., Moussaoui M., Jami M., Naji H. and Bouzidi M., Double MRT thermal lattice Boltzmann method for simulating convective flows, Physical Letters A, 374, pp. 3499-3507, 2010.

Peng Y., Shu C. and Chew Y., Simplified thermal lattice Boltzmann model for incompressible thermal flows, Physical Review E, 68, pp. 026701, 2003.

Succi S., Lattice Boltzmann Method for Fluid Dynamics and Beyond, Oxford University Press, 2001.

Wang J., Wang D., Lallemand P. and Luo L., Lattice Boltzmann simulations of thermal convective flows in two dimensions, Computers and Mathematics with Applications, 65, pp. 262-286, 2013.

Table 1 Rayleigh numbers and Prandtl numbers in four cases

|      | Case 1          | Case 2          | Case 3          | Case 4          |
|------|-----------------|-----------------|-----------------|-----------------|
| $Ra$ | $1\times10^4$   | $5\times10^4$   | $5\times10^4$   | $1\times10^5$   |
| $Pr$ | 0.01            | 0.01            | 0.005           | 0.01            |

# Figure Captions

Fig. 1 Natural convection model
Fig. 2 Nine directions in D2Q9 model
Fig. 3 Five directions in D2Q5 model
Fig. 4 Case 1: average Nusselt number
Fig. 5 Case 1: streamlines
Fig. 6 Case 1: temperature field
Fig. 7 Case 1: local Nusselt number along the heat wall
Fig. 8 Case 2: average Nusselt number tendency
Fig. 9 Case 2: streamlines
Fig. 10 Case 2: temperature field
Fig. 11 Case 3: average Nusselt number tendency
Fig. 12 Case 3: streamlines
Fig. 13 Case 3: temperature field
Fig. 14 Case 4: average Nusselt number tendency
Fig. 15 Case 4: streamlines
Fig. 16 Case 4: temperature field

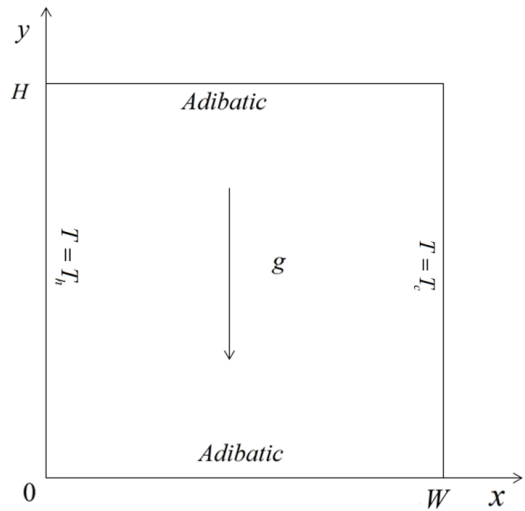

Fig. 1

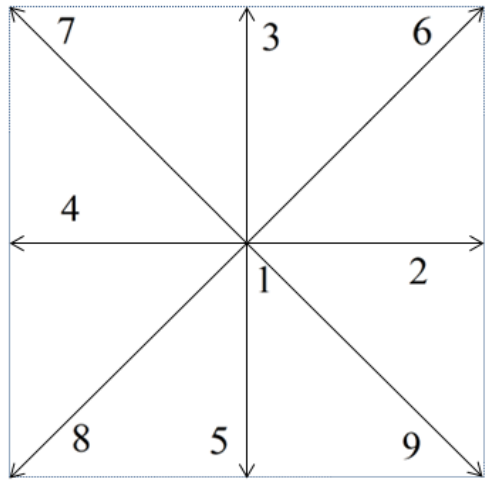

Fig. 2

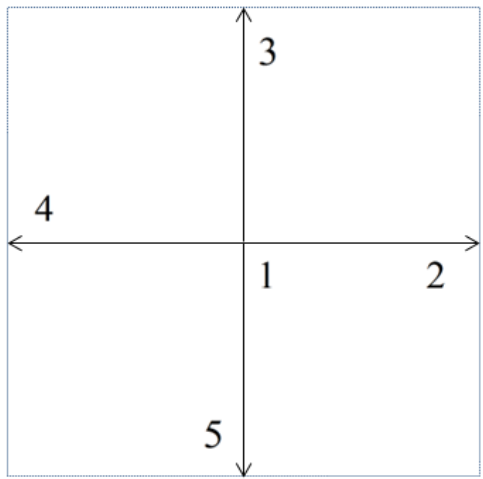

Fig. 3

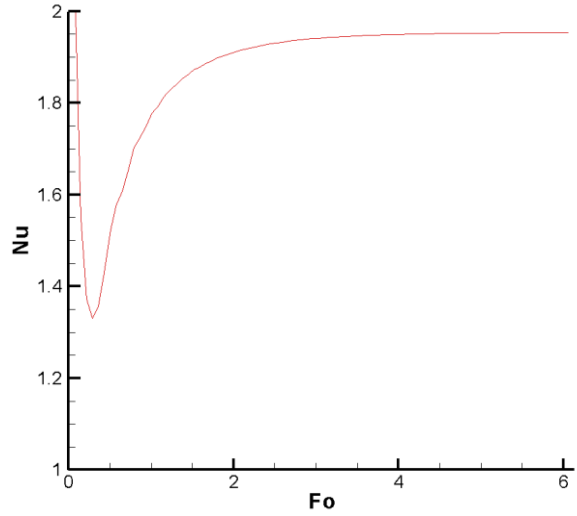

Fig. 4

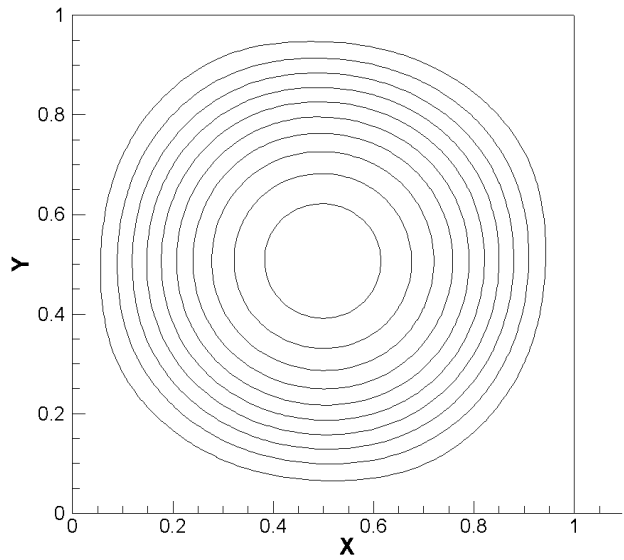

Fig. 5

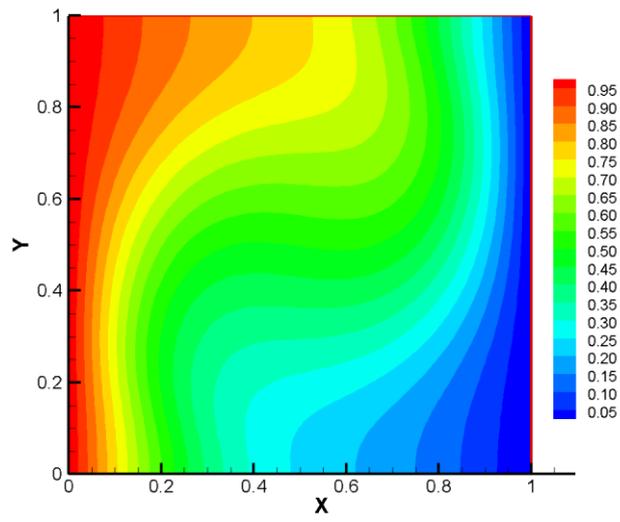

Fig. 6

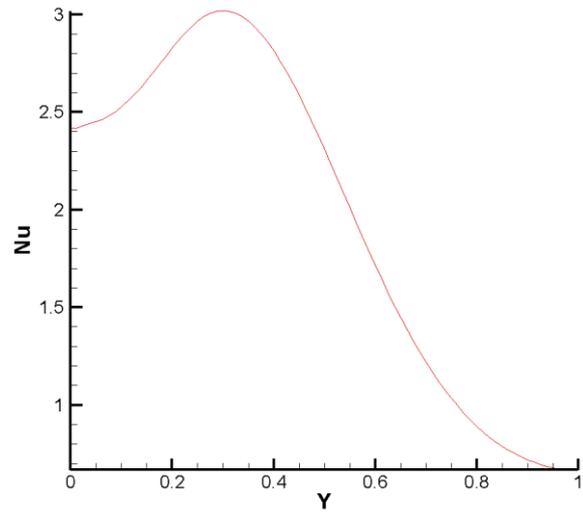

Fig. 7

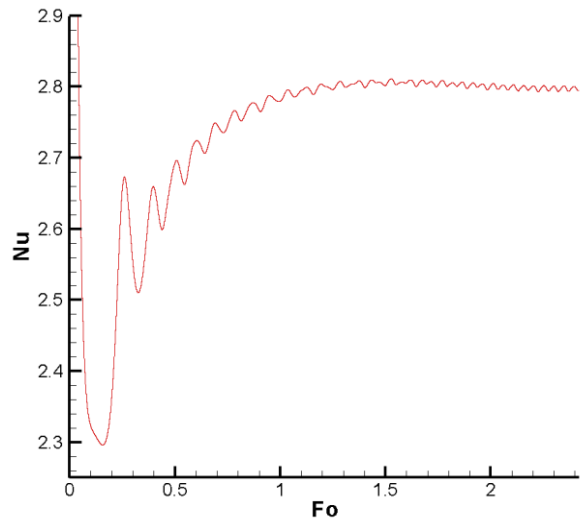

Fig. 8

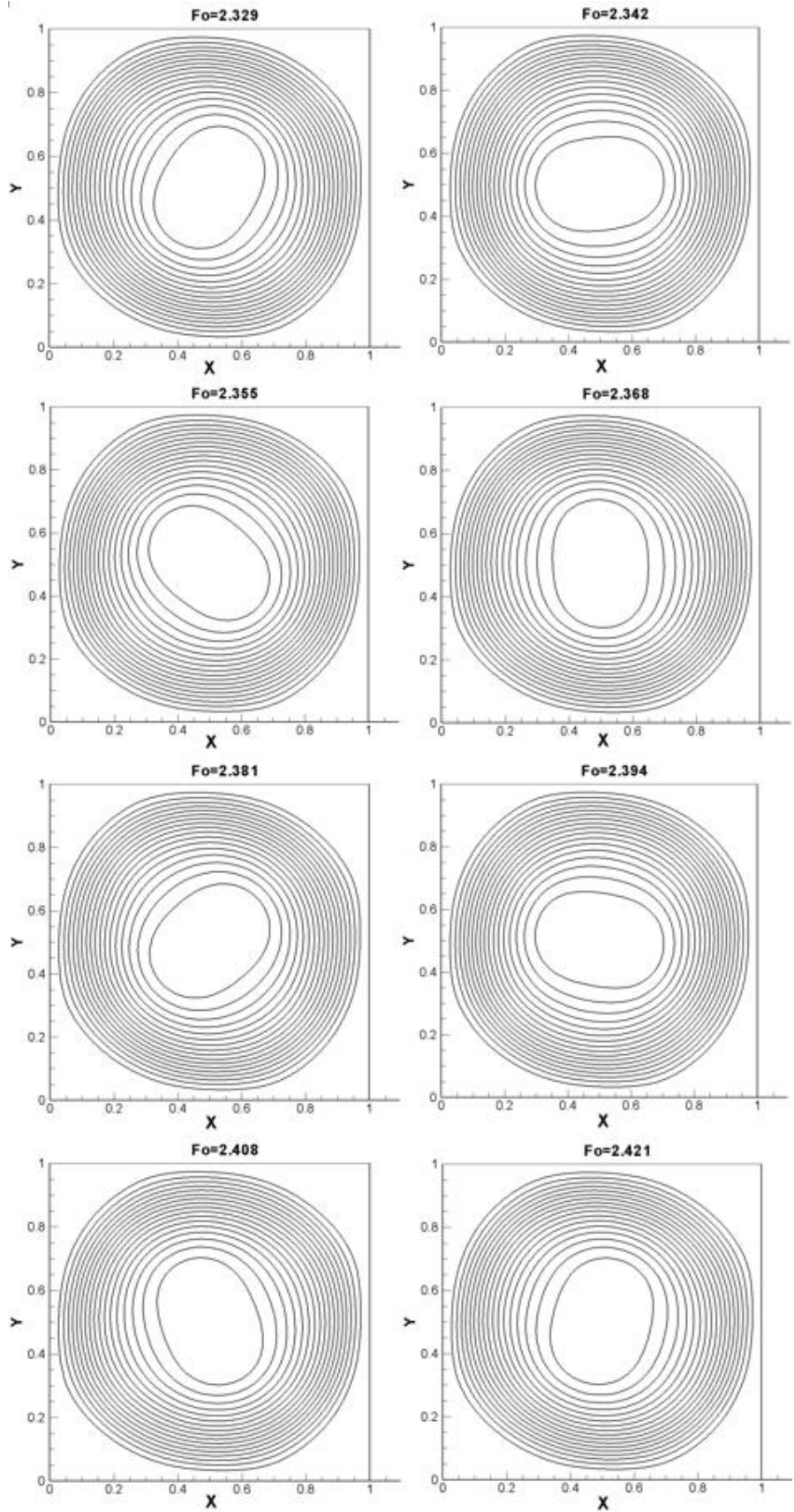

Fig. 9

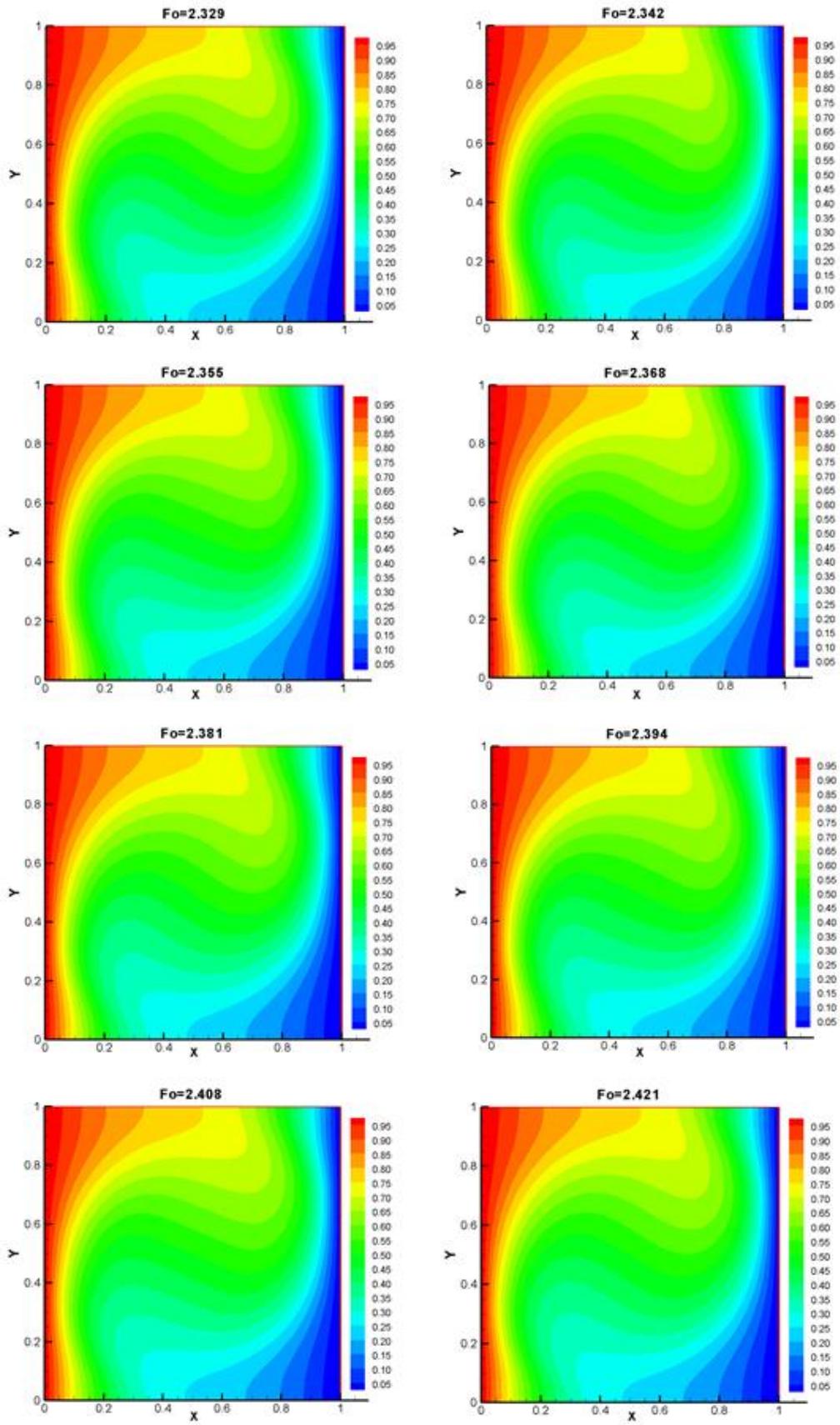

Fig. 10

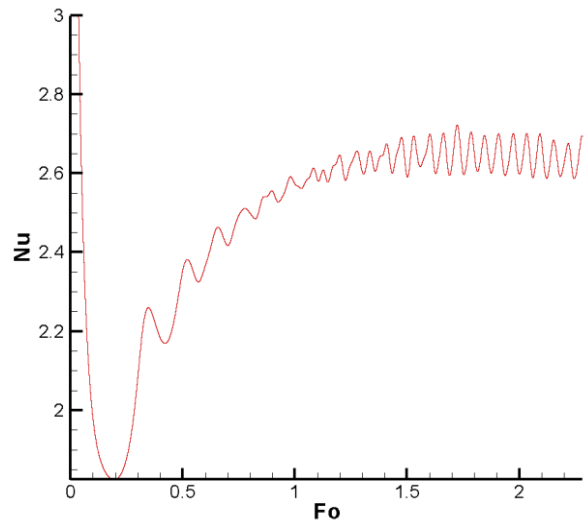
Fig. 11

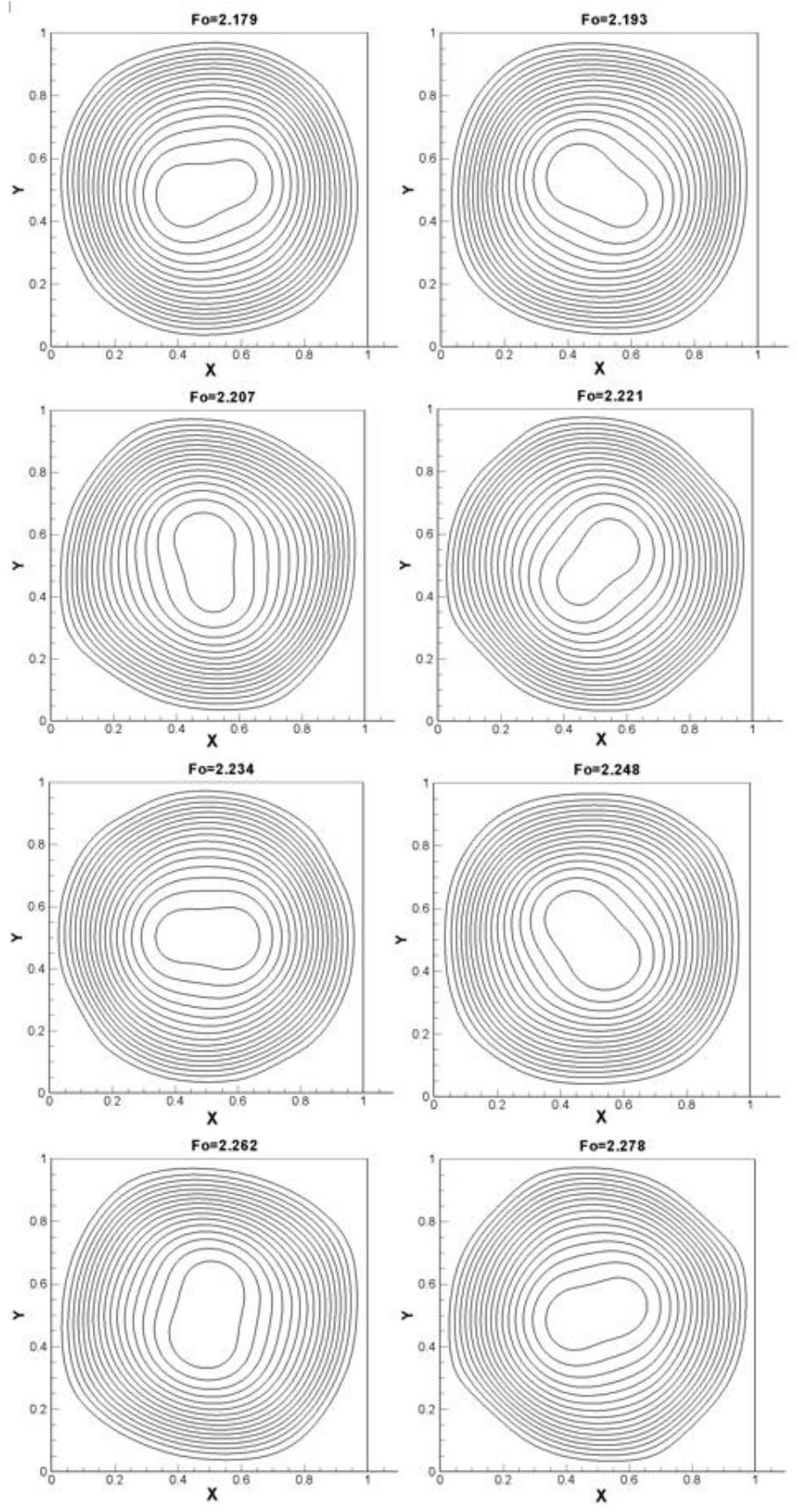

Fig. 12

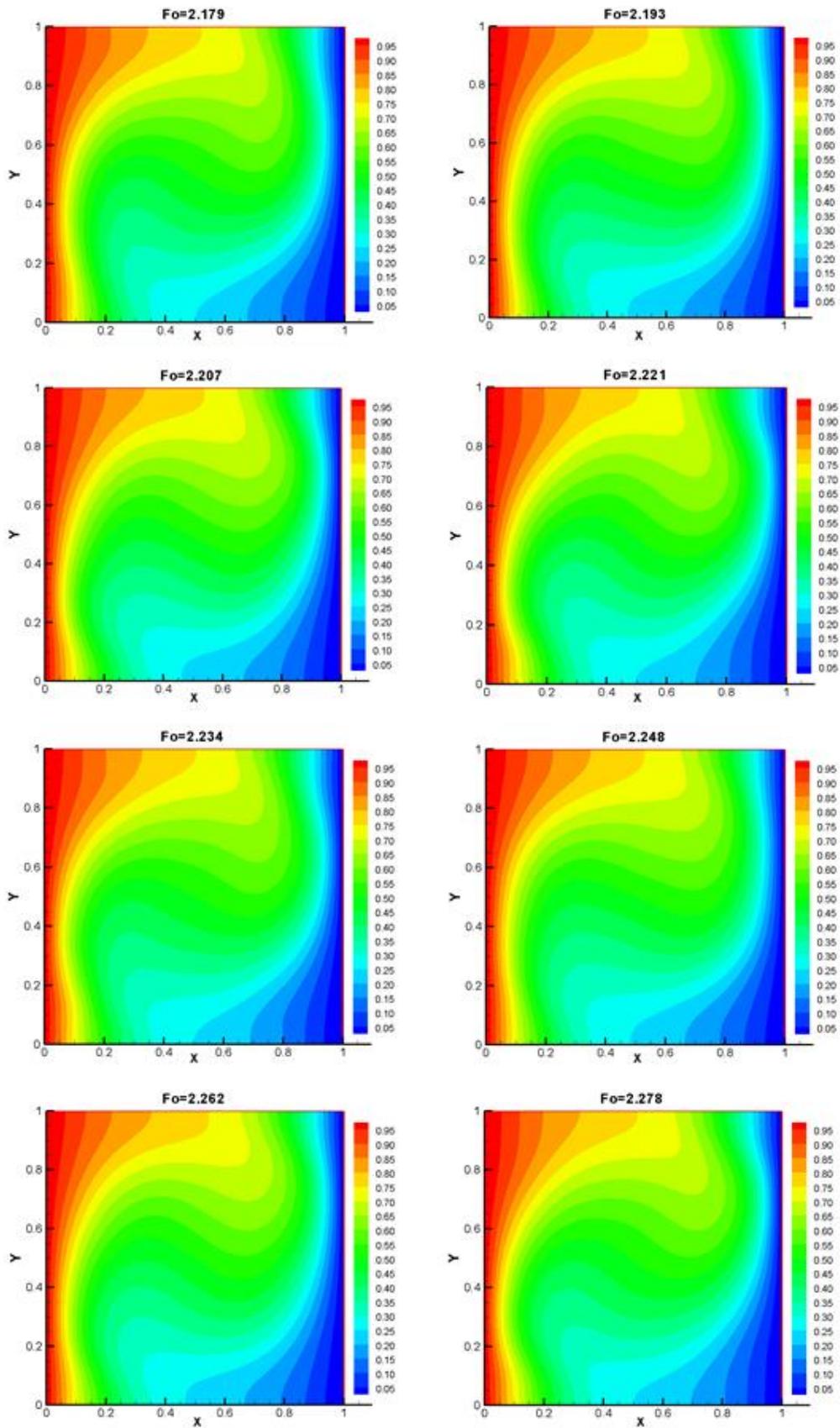

Fig. 13

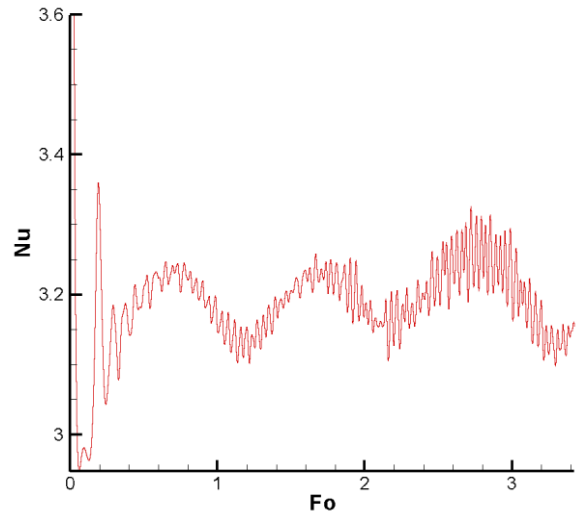
Fig. 14

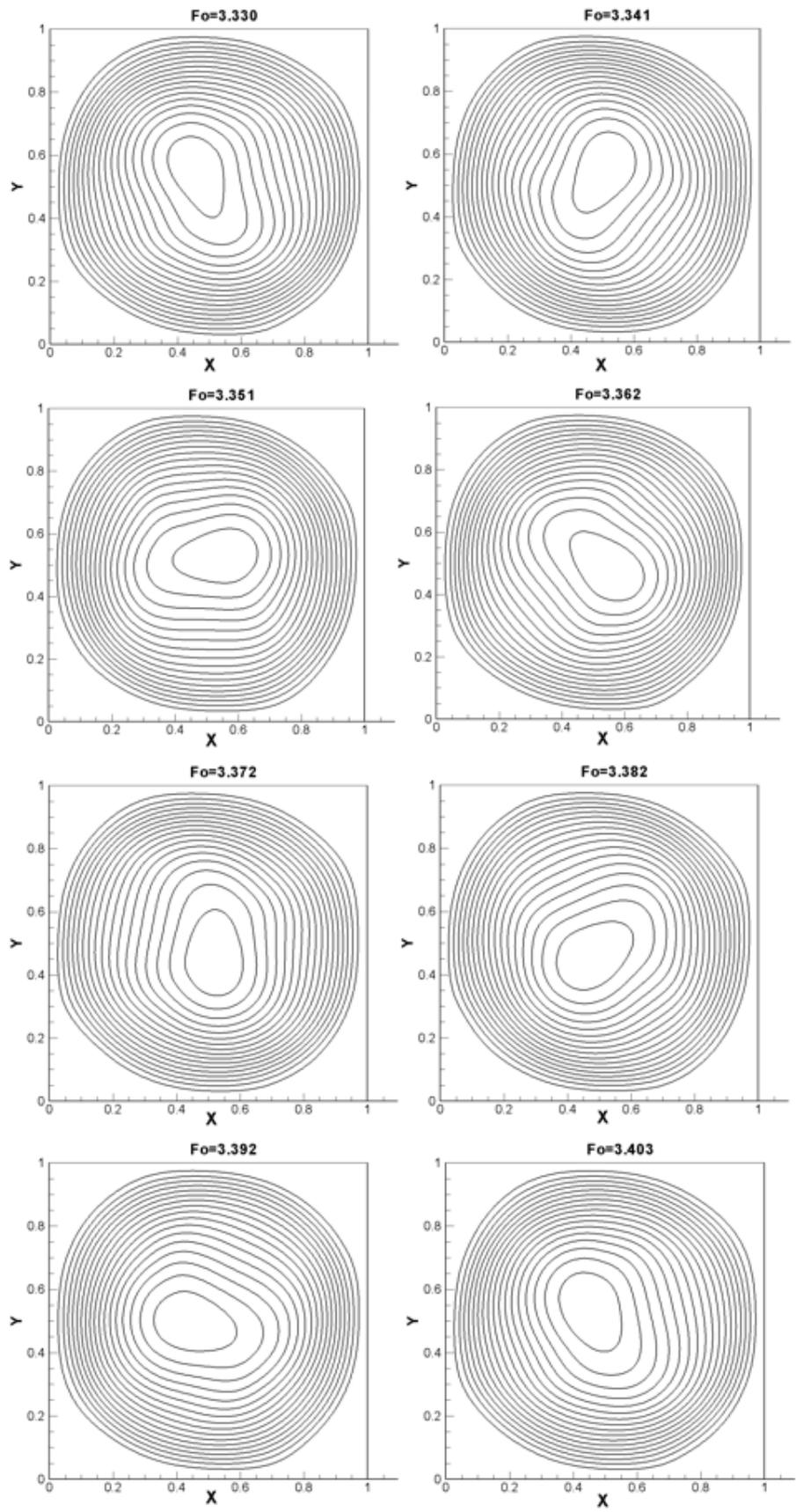

Fig. 15

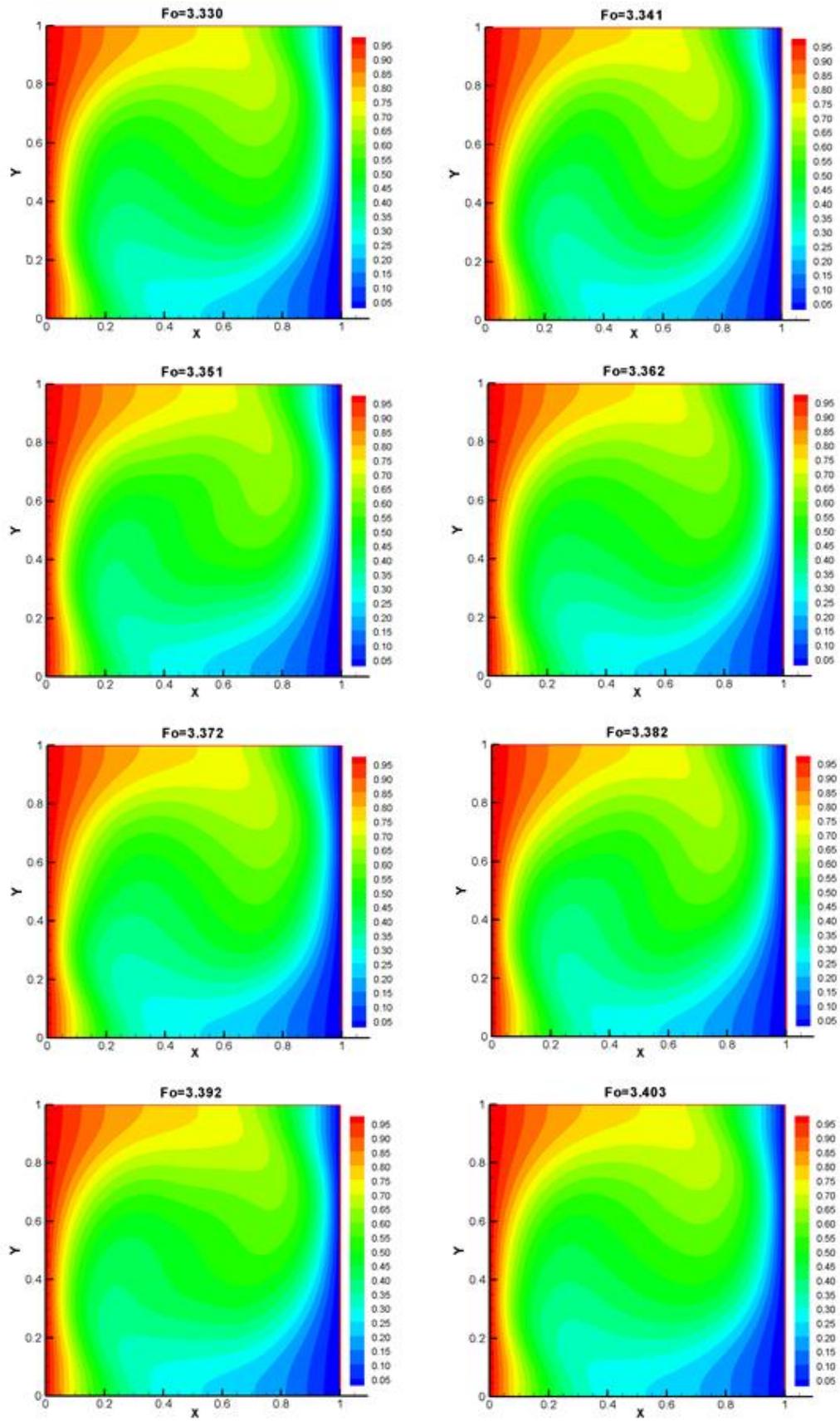

Fig. 16